\documentclass[aps,PRB,twocolumn,superscriptaddress,preprintnumbers]{revtex4-2}
\usepackage[T1]{fontenc}
\usepackage{times}
\usepackage{graphicx,color}
\usepackage{amsfonts,amsmath,amssymb,amsbsy}
\usepackage[colorlinks=true, linkcolor=blue, citecolor=blue, urlcolor=blue]{hyperref}
\usepackage{float}

\begin{document}

\title{Robustness of Majorana edge states of short-length Kitaev chains
coupled with environment}
\author{Motohiko Ezawa}
\affiliation{Department of Applied Physics, University of Tokyo, Hongo 7-3-1, 113-8656,
Japan}

\begin{abstract}
Recently, the two-site Kitaev model hosting Majorana edge states was
experimentally realized based on double quantum dots. In this context, we
construct two-band effective models describing Majorana edge states of a
finite-length Kitaev chain by using the isospectral matrix reduction method.
We analytically estimate the robustness of Majorana edge states as a
function of model parameters. We also study effects of coupling to an
environment based on non-Hermitian Hamiltonians derived from the Lindblad
equation. We study three types of dissipation, a local one, an adjacent one
and a global one. It is found that the Majorana zero-energy edge states
acquire nonzero energy such as $E\propto \pm \left( i\gamma \right) ^{L}$
for the local dissipation, where $\gamma $ is the magnitude of the
dissipation and $L$ is the length of the chain. On the other hand, the
Majorana zero-energy edge states acquire nonzero energy such as $E\propto
\pm i\gamma $ irrespective of the length $L$ for the global dissipation.
Hence, the Majorana edge states are robust against the local dissipation but
not against the global one. Our results will be useful for future studies on
Majorana edge states based on quantum dots.
\end{abstract}

\maketitle

\section{Introduction}

Majorana fermions are key for topological quantum computation\cite%
{Brav2,Ivanov,KitaevTQC,DasTQC,TQC,EzawaTQC}. The Majorana edge states form
a nonlocal qubit, which is robust for local perturbation. Thus, the qubit
based on Majorana fermions will resolve the problem of decoherence in
quantum computation. Majorana fermions are materialized in topological
superconductors\cite{Qi,Alicea,Sato,AliceaBraid}. A simplest model of a
topological superconductor hosting Majorana fermions is the Kitaev chain\cite%
{Kitaev01}. Despite the simpleness of the model, it is hard to materialize
it because it is hard to realize the $p$-wave superconducting order on the
lattice. Recently, the two-site Kitaev chain was experimentally realized in
double quantum dots\cite{Dvir}. In addition, the three-site Kitaev chain was
also experimentally realized in a nanowire device\cite{Bordin} \ It evokes
studies on the Minimal Kitaev chain based on double quantum dots\cite%
{Tsin,LiuB,Koch,TsinRev,Pino,Samu} and short-length quantum dots\cite%
{Mohse,Sout,Mile}. However, the Majorana edge states in the two-site Kitaev
chain is not topologically protected. Indeed, we need a precise tuning of
the model parameters so that the Majorana edge states exist exactly at the
zero energy. On the other hand, the Majorana edge states are robust if the
length of the Kitaev chain is long enough. It is hard to increase the number
of quantum dots in the current technology. There are studies on the
condition for the existence of the Majorana edge states to have exact zero
energy\cite{Kao,Hedge,Zvy,ZengC,Leum}. However, there is no construction of
the two-band effective model based on Majorana edge states for a
finite-length chain. It is intriguing if we can estimate the robustness of
the Majorana edge states for a finite-length Kitaev chain. It is discussed%
\cite{Kitaev01} that the robustness of the Majorana zero-energy states
increases exponentially as a function of the length of the Kitaev chain.

The platform of the Majorana fermions such as a quantum dot system has an
interaction with another system such as a substrate. In general, the
coupling to bath makes the system an open quantum system. It is commonly
analyzed based on the Lindblad equation\cite{Lind}. The short-time dynamics
is well described by a non-Hermitian Hamiltonian derived from the Lindblad
equation\cite{Lind}. A Kitaev chain with loss and gain is studied in the
context of non-Hermitian Hamiltonian\cite%
{Yuce,Zeng,Kawabata,KawabataNC,SatoX,Shibata,EzawaMajo,Zhao,Lieu}.

\begin{figure}[t]
\centerline{\includegraphics[width=0.48\textwidth]{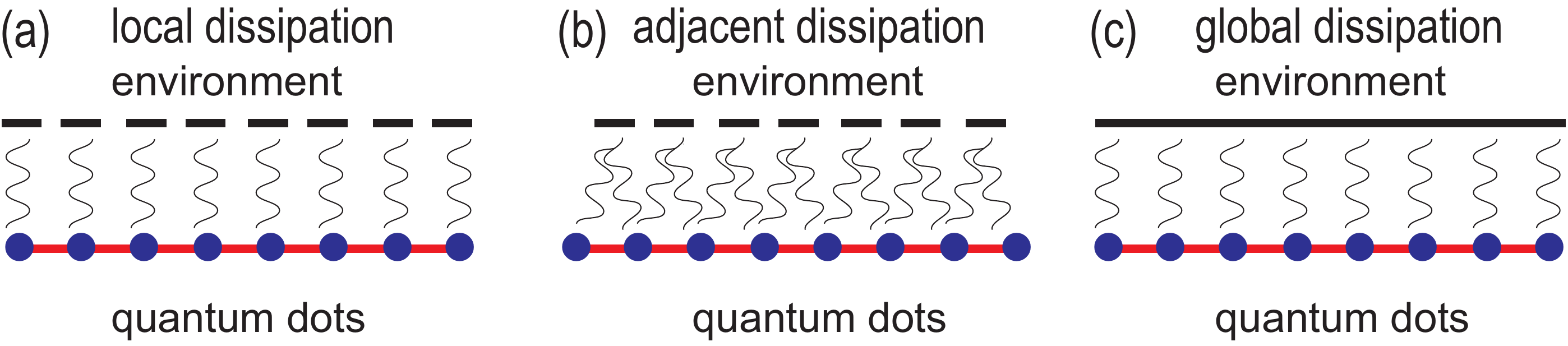}}
\caption{Illustration of (a) local dissipation, where a quantum dot and an
environment interact independently in each single quantum dot, (b) adjacent
dissipation, where two quantum dots and an environment interact in
nearest-neighbor sites, and (c) global dissipation, where every quantum dots
and an environment interact coherently.}
\label{FigIllust}
\end{figure}

In this paper, we construct effective two-band models for the Majorana edge
states by using the isospectral matrix reduction method\cite{Eek,Ront}. In
addition, we study the effect of the coupling between an environment based
on the non-Hermitian Hamiltonian formalism. We study three cases of
dissipation. The first one is a local dissipation, where there are hoppings
within a single site via an environment as in Fig.\ref{FigIllust}(a). The
second one is an adjacent dissipation, where there are hopping between
nearest-neighbor sites via an environment as in Fig.\ref{FigIllust}(b). The
last one is a global dissipation, where the every site is coherently coupled
via an environment as in Fig.\ref{FigIllust}(c). We find that the Majorana
edge states are robust against the local dissipation because $E\propto \pm
\left( i\gamma \right) ^{L}$ but not against the global dissipation because $%
E\propto \pm i\gamma $\ irrespective of the length $L$, when  the amplitude
of the dissipation $\gamma $ is small enough.

\begin{figure*}[t]
\centerline{\includegraphics[width=0.88\textwidth]{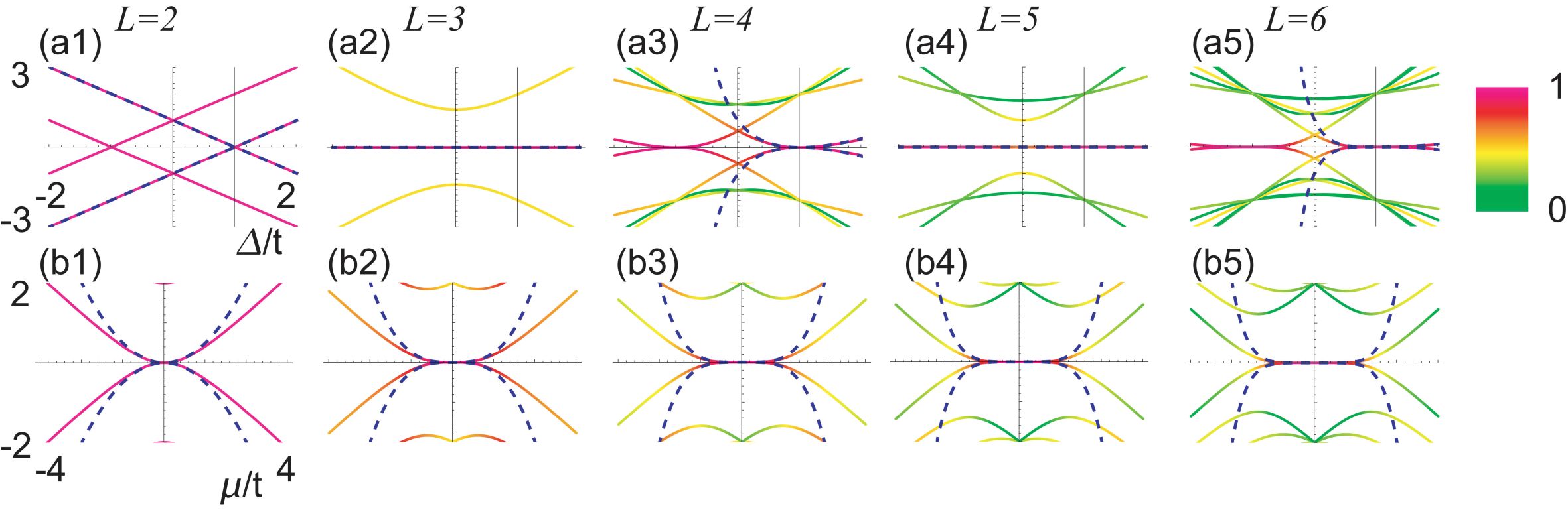}}
\caption{(a1)$\sim $(a4) The energy $E/t$ as a function of $\Delta /t$. We
have set $\protect\mu =0$. (b1)$\sim $(b4) The energy as a function of $%
\protect\mu /t$. We have set $\Delta =t$. Blue dotted curves are the
energy derived from the effective two-band model. (a1) and (b1) $L=2$, (a2) and (b2) $%
L=3$, (a3) and (b3) $L=4$, (a4) and (b4) $L=5$ and (a5) and (b5) $L=6$. (c)
Color palette indicating the amplitude at the edge sites, where red color
indicates the edge states and the green color indicates the bulk states. }
\label{FigKitaev}
\end{figure*}

\section{Kitaev chain}

The Kitaev p-wave superconductor model is defined on the 1D lattice as\cite%
{Kitaev01,Alicea}%
\begin{align}
\hat{H}& =-\mu \sum_{x=1}^{L-1}c_{x}^{\dagger }c_{x}-t\sum_{x=1}^{L-1}\left(
c_{x}^{\dagger }c_{x+1}+c_{x+1}^{\dagger }c_{x}\right)  \notag \\
& -\sum_{x=1}^{L-1}\left( \Delta c_{x}c_{x+1}+\Delta c_{x+1}^{\dagger
}c_{x}^{\dagger }\right) ,  \label{KitaevH}
\end{align}%
where $\mu $ is the chemical potential, $t>0$ is the nearest-neighbor
hopping strength, $\Delta >0$ is the $p$-wave pairing amplitude of the
superconductor, and $L$\ is the length of the chain.

\subsection{Majorana representation}

The system is topological for $\left\vert \mu \right\vert <2t$, where
Majorana edge states emerge at both the edges of the chain. We rewrite the
fermion operator in terms of two Majorana operators as 
\begin{equation}
c_{x}=\left( \frac{\gamma _{B,x}+i\gamma _{A,x}}{2}\right) ,\quad
c_{x}^{\dagger }=\frac{\gamma _{B,x}-i\gamma _{A,x}}{2},
\end{equation}%
where these Majorana operators satisfy 
\begin{equation}
\gamma _{\alpha ,x}=\gamma _{\alpha ,x}^{\dagger },\quad \left\{ \gamma
_{\alpha ,x},\gamma _{\alpha ^{\prime },x^{\prime }}\right\} =2\delta
_{\alpha \alpha ^{\prime }}\delta _{xx^{\prime }}
\end{equation}%
with $\alpha =A,B$. The Hamiltonian is rewritten in terms of Majorana
operators as%
\begin{align}
\hat{H}& =-\frac{\mu }{2}\sum_{x=1}^{L-1}\left( 1+i\gamma _{B,x}\gamma
_{A,x}\right)  \notag \\
& -i\sum_{x=1}^{L-1}\left[ \left( \Delta +t\right) \gamma _{B,x}\gamma
_{A,x+1}+\left( \Delta -t\right) \gamma _{A,x}\gamma _{B,x+1}\right] .
\end{align}

When $\mu =0$ and $t=\Delta \neq 0$, where the system is topological, the
Hamiltonian is simplified as%
\begin{equation}
\hat{H}=-2it\sum_{x=1}^{L-1}\gamma _{B,x}\gamma
_{A,x+1}=4t\sum_{x=1}^{L-1}\left( d_{x}^{\dagger }d_{x}-\frac{1}{2}\right) ,
\end{equation}%
where 
\begin{equation}
d_{x}=\frac{1}{2}\left( \gamma _{A,x+1}+i\gamma _{B,x}\right) ,\quad
d_{x}^{\dagger }=\frac{1}{2}\left( \gamma _{A,x+1}-i\gamma _{B,x}\right) .
\end{equation}%
The system is exactly solvable. The ground states are given by $%
d_{x}|0\rangle _{d}=0$ with the energy $-2t$, whose excited states are $%
|1\rangle _{d}=d_{x}^{\dagger }|0\rangle _{d}$ with the energy $2t$. They
constitute the bulk band. Apart from them, because this Hamiltonian does not
contain $\gamma _{A,1}$ and $\gamma _{B,N}$, there exist two Majorana states
perfectly localized at the two edge sites and exactly at the zero energy. A
non-local fermion operator is defined from them as 
\begin{equation}
f=\frac{1}{2}\left( \gamma _{A,1}+i\gamma _{B,N}\right) .
\end{equation}%
A qubit ($\left\vert 0\right\rangle _{\text{qubit}},\left\vert
1\right\rangle _{\text{qubit}}$) is constructed such as $f\left\vert
0\right\rangle _{\text{qubit}}=0$ and $\left\vert 1\right\rangle _{\text{%
qubit}}=f^{\dagger }\left\vert 0\right\rangle _{\text{qubit}}$. It is
interesting to note that the Majorana chain even with $L=2$ is possible to
support a qubit.

We show the energy spectrum of (\ref{KitaevH}) as a function of $\Delta /t$\
for $L=2,3,4,5,6$\ in Fig.\ref{FigKitaev}(a1)$\sim $(a5) and that as a
function of $\mu /t$\ in Fig.\ref{FigKitaev}(b1)$\sim $(b5). Especially, the
energy has a linear dependence $E=\left\vert \Delta -t\right\vert $ for the
two-site Kitaev chain with $\mu =0$ as shown in Fig.\ref{FigKitaev}(a1). On
the other hand, the energy has a parabolic dependence $E\propto \mu ^{2}$
for the two-site Kitaev chain with $\Delta =t$ as shown in Fig.\ref%
{FigKitaev}(b1).

The real part of the energy is exactly zero for odd length model as shown in
Fig.\ref{FigKitaev}(a2) and (a4). We will analytically verify these results
by deriving an effective two-band model in the following.

\begin{figure*}[t]
\centerline{\includegraphics[width=0.88\textwidth]{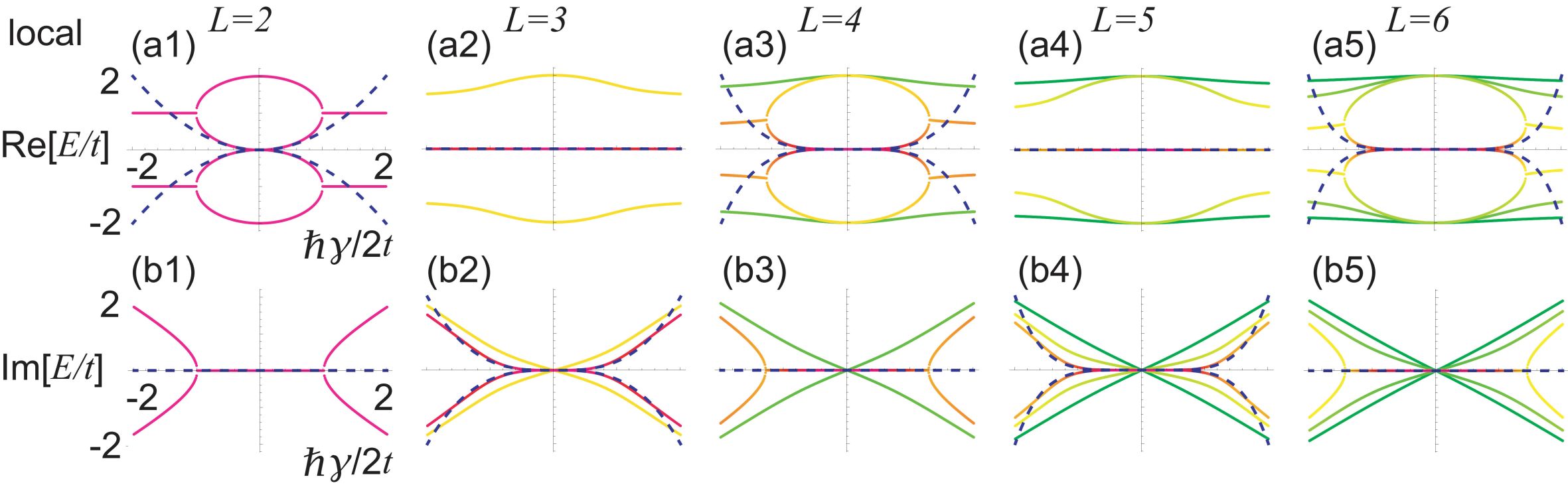}}
\caption{(a1)$\sim $(a4) real part and (b1)$\sim $(b4) imaginary part of
energy as a function of the local dissipation $\hbar \protect\gamma /2t$.
(a1) and (b1) $L=2$, (a2) and (b2) $L=3$, (a3) and (b3) $L=4$, (a4) and (b4) 
$L=5$ and (a5) and (b5) $L=6$. We have set $\Delta =t$ and $\protect\mu =0$. See the color palette in Fig.\ref{FigKitaev}.}
\label{FigA}
\end{figure*}

\begin{figure*}[t]
\centerline{\includegraphics[width=0.88\textwidth]{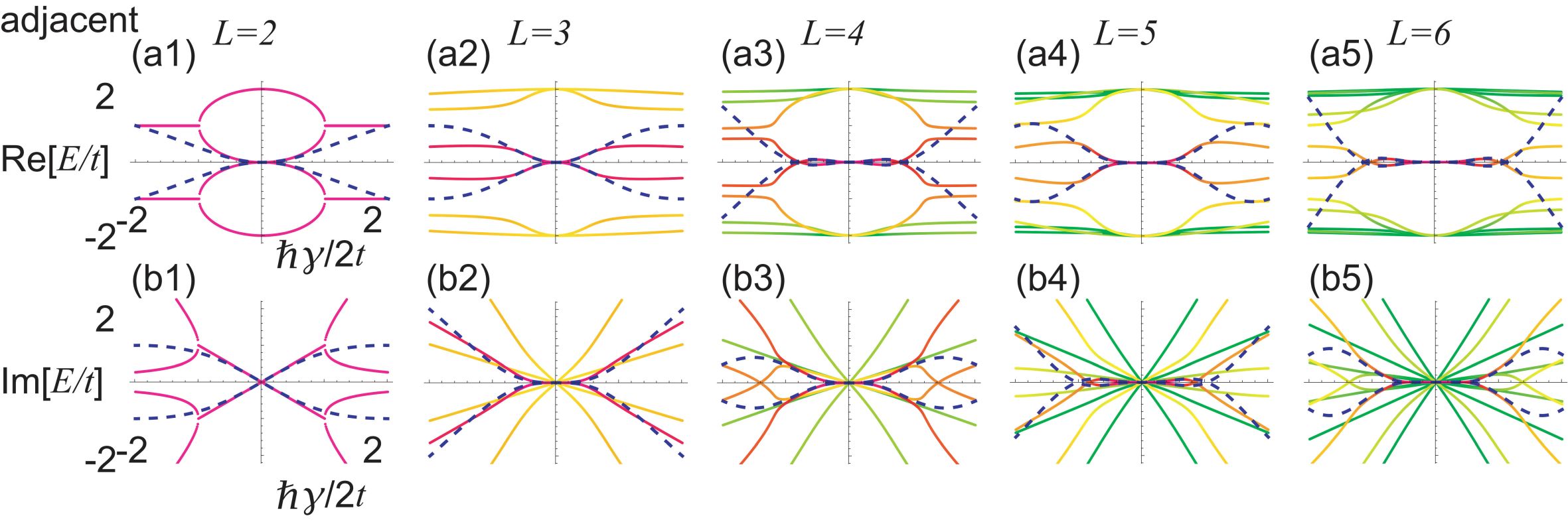}}
\caption{(a1)$\sim $(a4) real part and (b1)$\sim $(b4) imaginary part of
energy as a function of the adjacent dissipation $\hbar \protect\gamma /2t$.
The horizontal axis is $\protect\eta _{\text{adj}}$. (a1) and (b1) $L=2$,
(a2) and (b2) $L=3$, (a3) and (b3) $L=4$, (a4) and (b4) $L=5$ and (a5) and
(b5) $L=6$.  We have set $\Delta =t$ and $\protect\mu =0$. See the color palette in Fig.\ref{FigKitaev}.}
\label{FigB}
\end{figure*}

\section{Open quantum system}

Effects of the coupling between the system and an environment are described
by the Lindblad equation\cite{Lind} for the density matrix $\rho $ as%
\begin{equation}
\frac{d\rho }{dt}=-\frac{i}{\hbar }\left[ \hat{H},\rho \right] +\left(
\sum_{\alpha }L_{\alpha }\rho L_{\alpha }^{\dagger }-\frac{1}{2}\left\{
L_{\alpha }^{\dagger }L_{\alpha },\rho \right\} \right) ,
\end{equation}%
where $L$ is the Lindblad operator describing the dissipation. This equation
is rewritten in the form of%
\begin{equation}
\frac{d\rho }{dt}=-\frac{i}{\hbar }\left( \hat{H}_{\text{eff}}\rho -\rho 
\hat{H}_{\text{eff}}^{\dagger }\right) +\sum_{\alpha }L_{\alpha }\rho
L_{\alpha }^{\dagger },
\end{equation}%
where $H_{\text{eff}}$ is a non-Hermitian effective Hamiltonian defined by $%
H_{\text{eff}}\equiv H+H_{\text{dissipation}}$ with the dissipation
Hamiltonian 
\begin{equation}
\hat{H}_{\text{dissipation}}\equiv -\frac{i\hbar }{2}L^{\dagger }L.
\end{equation}%
It describes a short-time dynamics\cite{Lind}. We consider three types of
dissipations as illustrated in Fig.\ref{FigIllust}.

\begin{figure*}[t]
\centerline{\includegraphics[width=0.88\textwidth]{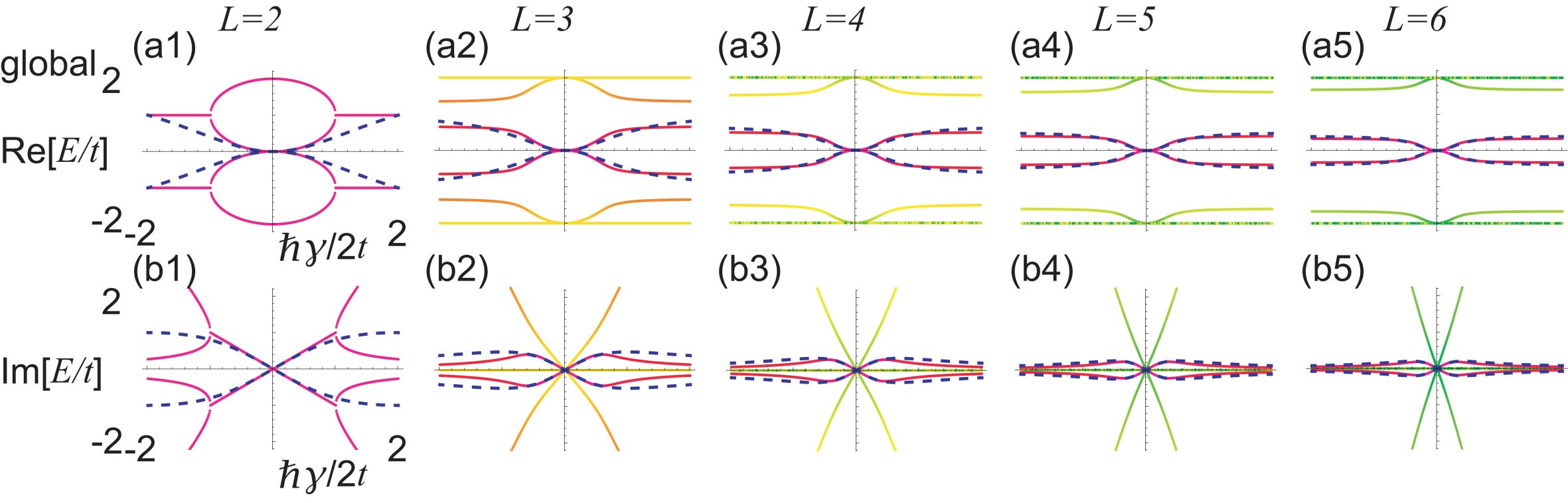}}
\caption{(a1)$\sim $(a4) real part and (b1)$\sim $(b4) imaginary part of
energy as a function of the global dissipation $\hbar \protect\gamma /2t$.
The horizontal axis is $\protect\eta _{\text{global}}$. (a1) and (b1) $L=2$,
(a2) and (b2) $L=3$, (a3) and (b3) $L=4$, (a4) and (b4) $L=5$ and (a5) and
(b5) $L=6$. \ We have set $\Delta =t$ and $\protect\mu =0$. See the color palette in Fig.\ref{FigKitaev}.}
\label{FigC}
\end{figure*}

\subsection{Local dissipation}

First, we study the local dissipation\cite{Lieu}, where the Lindblad
operators are given by 
\begin{equation}
L_{x}^{-}=\sqrt{\gamma _{-}}c_{x},\quad L_{x}^{+}=\sqrt{\gamma _{+}}%
c_{x}^{\dagger },
\end{equation}%
where $\gamma _{\pm }$ represent the dissipation. They describes the effect
that the particle is coming in and out of a single site. The corresponding
dissipation Hamiltonian reads%
\begin{equation}
\hat{H}_{\text{dissipation}}=-\frac{i\hbar }{2}\sum_{x=1}^{L}\left[ \gamma
c_{x}^{\dagger }c_{x}+\gamma _{+}\right] ,
\end{equation}%
where $\gamma \equiv \gamma _{-}-\gamma _{+}$. By introducing a complex
chemical potential%
\begin{equation}
\tilde{\mu}=\mu +\frac{i\hbar }{2}\gamma ,  \label{LocalMu}
\end{equation}%
the effect of the local dissipation is fully taken into account. We show the
energy spectrum as a function of $\gamma $ in Fig.\ref{FigA}. The real part
of the energy for odd length is exactly zero as shown in Fig.\ref{FigA}(a2)
and (a4). On the other hand, the imaginary part of the energy for even
length is exactly zero if the dissipation is smaller than a certain critical
value $\left\vert \gamma \right\vert <\left\vert \gamma _{\text{critical}%
}\right\vert $. The flat region of the zero-energy Majorana edge states is
expanded for longer chains. We derive these properties based on an effective
model with the use of the isospectral matrix reduction method in Sec.\ref%
{SecA}.

\subsection{Adjacent dissipation}

Next, we study the adjacent dissipation\cite{Diehl}, where the Lindblad
operators are given by 
\begin{equation}
L_{x}^{-}=\sqrt{\gamma _{-}/2}\left( c_{x}+c_{x+1}\right) ,\quad L_{j}^{+}=%
\sqrt{\gamma _{+}/2}\left( c_{x}^{\dagger }+c_{x+1}^{\dagger }\right) .
\end{equation}%
They describe the effect that the particle can hop between nearest-neighbor
sites via an environment. The corresponding dissipation Hamiltonian reads%
\begin{align}
& \hat{H}_{\text{dissipation}}-\frac{\gamma _{+}}{2}  \notag \\
& =-\frac{i\hbar }{2}\sum_{x=1}^{L-1}\frac{\gamma }{2}\left( c_{x}^{\dagger
}c_{x}+c_{x}^{\dagger }c_{x+1}+c_{x+1}^{\dagger }c_{x}+c_{x+1}^{\dagger
}c_{x+1}\right) .
\end{align}%
The effect of the adjacent dissipation is taken into account by considering
the complex chemical potential and the complex hopping defined by%
\begin{equation}
\tilde{\mu}=\mu +\frac{i\hbar }{2}\gamma ,\qquad \tilde{t}=t+\frac{i\hbar }{2%
}\gamma .  \label{AdjMu}
\end{equation}%
We show the energy spectrum as a function of $\gamma $ in Fig.\ref{FigB}.
The imaginary part of the energy acquire nonzero energy as a linear function
of $\gamma $ for $L=2$ as shown in Fig.\ref{FigB}(b1). This is because the
two Majorana edge states couple directly via the adjacent dissipation term.
It follows from Fig.\ref{FigB} that deviations from the zero energy of the
Majorana edge states becomes smaller for longer chains. Comparing Figs.\ref%
{FigA} and \ref{FigB}, the Majorana edge states are found to be more fragile
than the local dissipation. We derive these properties based on an effective
model with the use of the isospectral matrix reduction method in Sec.\ref%
{SecB}.

\subsection{Global dissipation}

Finally, we study the global dissipation, where the Lindblad operators are
given by \ 
\begin{equation}
L_{x}^{-}=\sqrt{\gamma _{-}}\sum_{x=1}^{L}c_{x},\quad L_{x}^{+}=\sqrt{\gamma
_{+}}\sum_{x=1}^{L}c_{x}^{\dagger }.
\end{equation}%
They describe the effect that all particles are equally coupled via an
environment. The corresponding dissipation Hamiltonian reads 
\begin{eqnarray}
\hat{H}_{\text{dissipation}} &=&-\frac{i\hbar }{2}\gamma \sum_{x,x^{\prime
}=1}^{L}\left( c_{x}^{\dagger }c_{x}+c_{x}^{\dagger }c_{x^{\prime
}}+c_{x^{\prime }}^{\dagger }c_{x}+c_{x^{\prime }}^{\dagger }c_{x^{\prime
}}\right)   \notag \\
&&+\gamma _{+},
\end{eqnarray}%
which is highly nonlocal. We show the energy spectrum as a function of $%
\gamma $ in Fig.\ref{FigC} for $L=2,3,4,5$. It is found from Fig.\ref{FigC}%
(b1)\symbol{126}(b5) that the imaginary part of the energy is linear as a
function of $\gamma $ irrespective of the length of the chain although the
slope becomes smaller for a longer chain. Hence, the Majorana edge states
are not robust in the presence of the global dissipation even for a long
Kitaev chain. It is natural because the global dissipation term couples the
Majorana edge states directly. We derive these properties based on an
effective model with the use of the isospectral matrix reduction method in
Sec.\ref{SecC}.

\section{Minimal Kitaev chain}

In the presence of the local dissipation or the adjacent dissipation, the
Hamiltonian of the two-site Kitaev chain reads 
\begin{equation}
\hat{H}=\left( 
\begin{array}{cccc}
c_{1}^{\dagger } & c_{2}^{\dagger } & c_{1} & c_{2}%
\end{array}%
\right) H\left( 
\begin{array}{c}
c_{1} \\ 
c_{2} \\ 
c_{1}^{\dagger } \\ 
c_{2}^{\dagger }%
\end{array}%
\right) ,
\end{equation}%
with%
\begin{equation}
H=\left( 
\begin{array}{cccc}
-\tilde{\mu} & -\tilde{t} & 0 & \Delta  \\ 
-\tilde{t} & -\tilde{\mu} & -\Delta  & 0 \\ 
0 & -\Delta  & \tilde{\mu} & \tilde{t} \\ 
\Delta  & 0 & \tilde{t} & \tilde{\mu}%
\end{array}%
\right) 
\end{equation}%
where we use the complex chemical potential $\tilde{\mu}$ in (\ref{LocalMu})
for the local dissipation and the complex chemical potential $\tilde{\mu}$
and the complex hopping $\tilde{t}$ in (\ref{AdjMu}) for the adjacent
dissipation. The energy spectrum is exactly obtained as 
\begin{equation}
E=\pm \tilde{t}\pm \sqrt{\Delta ^{2}+\tilde{\mu}^{2}}.
\end{equation}%
We note that the adjacent dissipation and the global dissipation is
identical for $L=2$.

\section{Isospectral matrix reduction method}

It is impossible to exactly diagonalize the Hamiltonian matrix hosting
Majorana edge states\ except for the exactly solvable parameter $\mu =0$, $%
\Delta =\pm t$ for the Kitaev chain for $L\geq 3$.

We derive an effective two-band model near the zero-energy describing the
Majorana edge states based on the isospectral matrix reduction method\cite%
{Eek,Ront}. We first diagonalize the Hamiltonian matrix for $\mu =0$, $%
\Delta =t$ as%
\begin{eqnarray}
H^{\prime } &\equiv &UHU^{-1}  \notag \\
&=&\text{diag.}\left\{ 0,0,-2t,-2t,\cdots ,-2t,2t,2t,\cdots ,2t\right\} .
\end{eqnarray}%
The first two zero-energy states correspond to the Majorana edge states.
Then we divide $H^{\prime }$ into the form of%
\begin{equation}
H^{\prime }=\left( 
\begin{array}{cc}
H_{1} & V \\ 
V^{\dagger } & H_{2}%
\end{array}%
\right) ,
\end{equation}%
where $H_{1}$ is the $2\times 2$ matrix and $H_{2}$ is the $(2L-2)\times
(2L-2)$ matrix. The eigenequation reads%
\begin{equation}
\left( 
\begin{array}{cc}
H_{1} & V \\ 
V^{\dagger } & H_{2}%
\end{array}%
\right) \left( 
\begin{array}{c}
\psi _{1} \\ 
\psi _{2}%
\end{array}%
\right) =E\left( 
\begin{array}{c}
\psi _{1} \\ 
\psi _{2}%
\end{array}%
\right) ,
\end{equation}%
from which we derive 
\begin{equation}
\psi _{2}=\left( E-H_{2}\right) ^{-1}V^{\dagger }\psi _{1}.
\end{equation}%
Then, we obtain a single nonlinear eigen equation for $\psi _{1}$ as 
\begin{equation}
\tilde{H}\left( E\right) \psi _{1}=E\psi _{1},
\end{equation}%
where 
\begin{equation}
\tilde{H}\left( E\right) =H_{1}+V\left( E-H_{2}\right) ^{-1}V^{\dagger }.
\end{equation}%
Exact solutions may be obtained by solving the nonlinear equation. However,
it is practically impossible to solve it because it is an algebraic equation
of the order of $E^{2L}$. Instead, we seek a solution in the vicinity of the
zero energy, where the Hamiltonian is well approximated by 
\begin{equation}
H_{\text{eff}}\equiv H_{1}-VH_{2}^{-1}V^{\dagger }.
\end{equation}%
The second term is written in the form of%
\begin{equation}
-VH_{2}^{-1}V^{\dagger }=F\sigma _{x}.
\end{equation}%
We explicitly determine $H_{\text{eff}}$ in what follows. 

\subsection{Hermitian model}

First, we derive an effective two-band model for the Hermitian system. We
find 
\begin{equation}
H_{1}=\left( t-\Delta \right) \sigma _{x}
\end{equation}%
for $L=2$ and $H_{1}=0$ for $L\geq 3$. On the other hand, we find%
\begin{equation}
F=\frac{1}{\left( \Delta +t\right) ^{L-1}}\sum_{m=0}^{\left\lfloor
L/2\right\rfloor }\left( 
\begin{array}{c}
L-m \\ 
m%
\end{array}%
\right) \mu ^{L-2m}\left( \Delta ^{2}-t^{2}\right) ^{2m}
\end{equation}%
for $L\geq 2$. The energy is given by 
\begin{equation}
E=\pm F.
\end{equation}

If $\mu =0$, we have $F=0$ for odd $L$, which well describes the energy near
the zero energy as shown in Fig.\ref{FigKitaev}(a2) and (a4). On the other
hand, we find 
\begin{equation}
F=-\frac{\left( \Delta -t\right) ^{L/2}}{\left( \Delta +t\right) ^{L/2+1}}%
\propto \left( \Delta -t\right) ^{L/2}
\end{equation}%
for even $L$, which well describes the energy near the zero energy as shown
in Fig.\ref{FigKitaev}(a1), (a3) and (a5). It is almost zero for small $%
\left\vert \Delta -t\right\vert $ and large $L$, where the Majorana edge
states are robust.

If $\Delta =t$, we have 
\begin{equation}
F=-\frac{\mu ^{L}}{\left( 2t\right) ^{L-1}}\propto \mu ^{L}.
\end{equation}%
It well fits the energy near the zero energy as shown in Fig.\ref{FigKitaev}%
(b1)$\sim $(b5). It is almost zero for small $\mu $ and large $L$, where the
Majorana edge states are robust.

\subsection{Local dissipation}

\label{SecA}

Next, we derive effective two-band models for the local dissipation with $%
\Delta =t$ and $\mu =0$. We find 
\begin{equation}
H_{\text{eff}}=-\frac{\left( i\gamma \right) ^{L}}{\left( 2t\right) ^{L-1}}
\end{equation}%
for $L\geq 2$. This formula well explains the fact that the real [imaginary]
part of the energy is zero for even [odd] $L$ as shown in Fig.\ref{FigA}(a2)
and (a4), [(b1) and (b3)]. Hence the Majorana edge states becomes robust for
a long chain.

\subsection{Adjacent dissipation}

\label{SecB}

We derive effective two-band models for the adjacent dissipation with $\mu =0
$. We find 
\begin{equation}
H_{1}=\left( t+i\gamma -\Delta \right) \sigma _{x}
\end{equation}%
for $L=2$ and $H_{1}=0$ for $L\geq 3$. On the othe hand, 
\begin{equation}
F=-\sum_{m=0}^{\left\lfloor L/2\right\rfloor }\left( 
\begin{array}{c}
L-m \\ 
m%
\end{array}%
\right) \frac{\left( i\gamma \right) ^{L-2m}}{\left( \Delta +t\right) ^{L-1}}%
\left( \Delta ^{2}-\left( t+i\gamma \right) ^{2}\right) ^{m}
\end{equation}%
for $L\geq 2$. For small $\gamma $ ($\left\vert \gamma /t\right\vert \ll 1$%
), they are explicitly given by%
\begin{equation}
\begin{array}{ll}
F=-i\gamma -\gamma ^{2}/2t+\cdots  & \text{for }L=2, \\ 
F=-3i\gamma ^{3}/4t^{2}+\gamma ^{2}/2t+\cdots  & \text{for }L=3, \\ 
F=-i\gamma ^{3}/t^{2}-\gamma ^{2}/2t+\cdots  & \text{for }L=4, \\ 
F=-i\gamma ^{3}/4t^{2}-\gamma ^{4}/4t^{3}+\cdots  & \text{for }L=5, \\ 
F=i\gamma ^{3}/4t^{2}+\gamma ^{4}/t^{3}+\cdots  & \text{for }L=6,%
\end{array}%
\end{equation}%
In general,%
\begin{equation}
F=ia\gamma ^{A}+b\gamma ^{B}+\cdots ,
\end{equation}%
with certain real numbers $a$ and $b$, where%
\begin{equation}
A=2\left\lfloor \frac{L+1}{4}\right\rfloor +1,\qquad B=2\left\lfloor \frac{%
L-1}{4}\right\rfloor +2.
\end{equation}%
The robustness of the Majorana edge modes is enhanced when the chain length
becomes longer. However, it is more fragile than the local dissipation.

\subsection{Global dissipation}

\label{SecC}

Finally, we derive effective two-band models for the global dissipation with 
$\Delta =t$ and $\mu =0$. We find%
\begin{equation}
H_{1}=i\gamma \sigma _{x},\quad F=\frac{\left( L-1\right) \gamma ^{2}}{%
2t+i\left( L-1\right) \gamma }\sigma _{x}
\end{equation}%
for $L\geq 2$. For small $\gamma $ ($\left\vert \gamma /t\right\vert \ll 1$%
), the Hamiltonian is approximated as%
\begin{equation}
H_{\text{eff}}\simeq \left( i\gamma +\frac{\left( L-1\right) \gamma ^{2}}{2t}%
\right) \sigma _{x}.  \label{H3g}
\end{equation}%
The real part of the energy is parabolic as a function of $\gamma $\
irrespective of the length $L$\ as shown in Fig.\ref{FigC}(a1)$\sim $(a5),
while the imaginary part of the energy is linear as a function of $\gamma $\
irrespective of the length $L$\ as shown in Fig.\ref{FigC}(b1)$\sim $(b5).
Hence, the robustness of the Majorana edge states is not enhanced even if we
use a chain with long length $L$\ in the presence of the global dissipation.

Actually, the energy spectrum in the vicinity of the zero energy is exactly
obtained without using the isospectral matrix reduction method at the point $%
\mu =0$ and $\Delta =t$, whose results reads%
\begin{equation}
F=i\frac{L}{2}\gamma +2t+i\sqrt{\left( \frac{L}{2}\gamma \right)
^{2}-2i\left( L-2\right) t\gamma -4t^{2}}.
\end{equation}%
It gives the same result as in (\ref{H3g}) for small $\gamma $ ($\left\vert
\gamma /t\right\vert \ll 1$).

\section{Discussion}

We have constructed two-band effective models describing the Majorana edge
states in the presence of three types of dissipation. We have found the
even-odd effect on the stability of the Majorana edge states. We have also
found that the robustness of the Majorana edge states is affected by the
type of dissipation, where the global dissipation is detrimental for the
stability of the Majorana edge states. The local dissipation may be realized
when the system couples with the substrate where the real coordinate is a
good quantum number. On the other hand, the global dissipation may be
realized when the system couples with the substrate where the momentum
coordinate is a good quantum number. Our results will be useful for
experimental realization of the Kitaev chain based on quantum dots.

This work is supported by CREST, JST (Grants No. JPMJCR20T2) and
Grants-in-Aid for Scientific Research from MEXT KAKENHI (Grant No.
23H00171). 


\end{document}